\documentclass[aps,prl,letterpaper,10pt,twocolumn,showpacs,superscriptaddress]{revtex4-1}

\usepackage{amsfonts,amssymb,amsmath,amsthm,graphicx}
\usepackage{url}
\usepackage{dsfont}
\usepackage{bbm}
\usepackage[usenames,dvipsnames]{xcolor}
\usepackage{nicefrac}
\usepackage{natbib}
\usepackage{setspace}
\usepackage{subfigure}
\usepackage{comment}
\usepackage{enumitem}

\usepackage{physics}
\usepackage{amsmath}
\usepackage{graphicx}
\usepackage[colorinlistoftodos]{todonotes}
\usepackage[colorlinks=true, allcolors=blue]{hyperref}

\usepackage{times,todonotes}
\usepackage[normalem]{ulem} 

\DeclareMathOperator{\sgn}{sgn}

\newcommand{\Id}{\mathds{1}}

\newcommand{\red}{\color{black}}

\newcommand{\beq}{\begin{equation}}
\newcommand{\eeq}{\end{equation}}

\begin{document}

\title{Postselection-loophole-free Bell violation with genuine time-bin entanglement}

\author{Francesco Vedovato}
\affiliation{Dipartimento di Ingegneria dell'Informazione, Universit\`a di Padova, via Gradenigo 6B, 35131 Padova, Italy}
\affiliation{Centro di Ateneo di Studi e Attivit\`a Spaziali ``G. Colombo", Universit\`a di Padova, via Venezia 15, 35131 Padova, Italy}
\author{Costantino Agnesi}
\author{Marco Tomasin} 
\author{Marco Avesani}
\affiliation{Dipartimento di Ingegneria dell'Informazione, Universit\`a di Padova, via Gradenigo 6B, 35131 Padova, Italy}
\author{Jan-\r{A}ke Larsson}
\affiliation{Institutionen f\"{o}r systemteknik, Link\"{o}ping Universitet,  581 83 Link\"{o}ping, Sweden}
\author{Giuseppe Vallone}
\author{Paolo Villoresi}
\affiliation{Dipartimento di Ingegneria dell'Informazione, Universit\`a di Padova, via Gradenigo 6B, 35131 Padova, Italy}
\affiliation{Istituto di Fotonica e Nanotecnologie, CNR, via Trasea 7, 35131 Padova, Italy}

\date{\today}

\begin{abstract}
Entanglement is an invaluable resource for fundamental tests of physics and the implementation of quantum information protocols such as device-independent secure communications. In particular, time-bin entanglement is widely exploited to reach these purposes both in free-space and optical fiber propagation,  due to  the robustness and simplicity of its implementation. However, all existing realizations of time-bin entanglement  suffer from an intrinsic postselection loophole, which undermines their usefulness. Here, we report the first experimental violation of Bell's inequality with ``genuine'' time-bin entanglement, free of the postselection loophole. We introduced a novel function of the interferometers at the two measurement stations, that operate as fast synchronized optical switches. This scheme allowed to obtain a postselection-loophole-free Bell violation of more than nine standard deviations. Since our scheme is fully implementable using standard fiber-based components and is compatible with modern integrated photonics, our results pave the way for the distribution of genuine time-bin entanglement over long distances.    
\end{abstract}

\maketitle

\begin{figure}[t!]
\centering\includegraphics[width=0.45\textwidth]{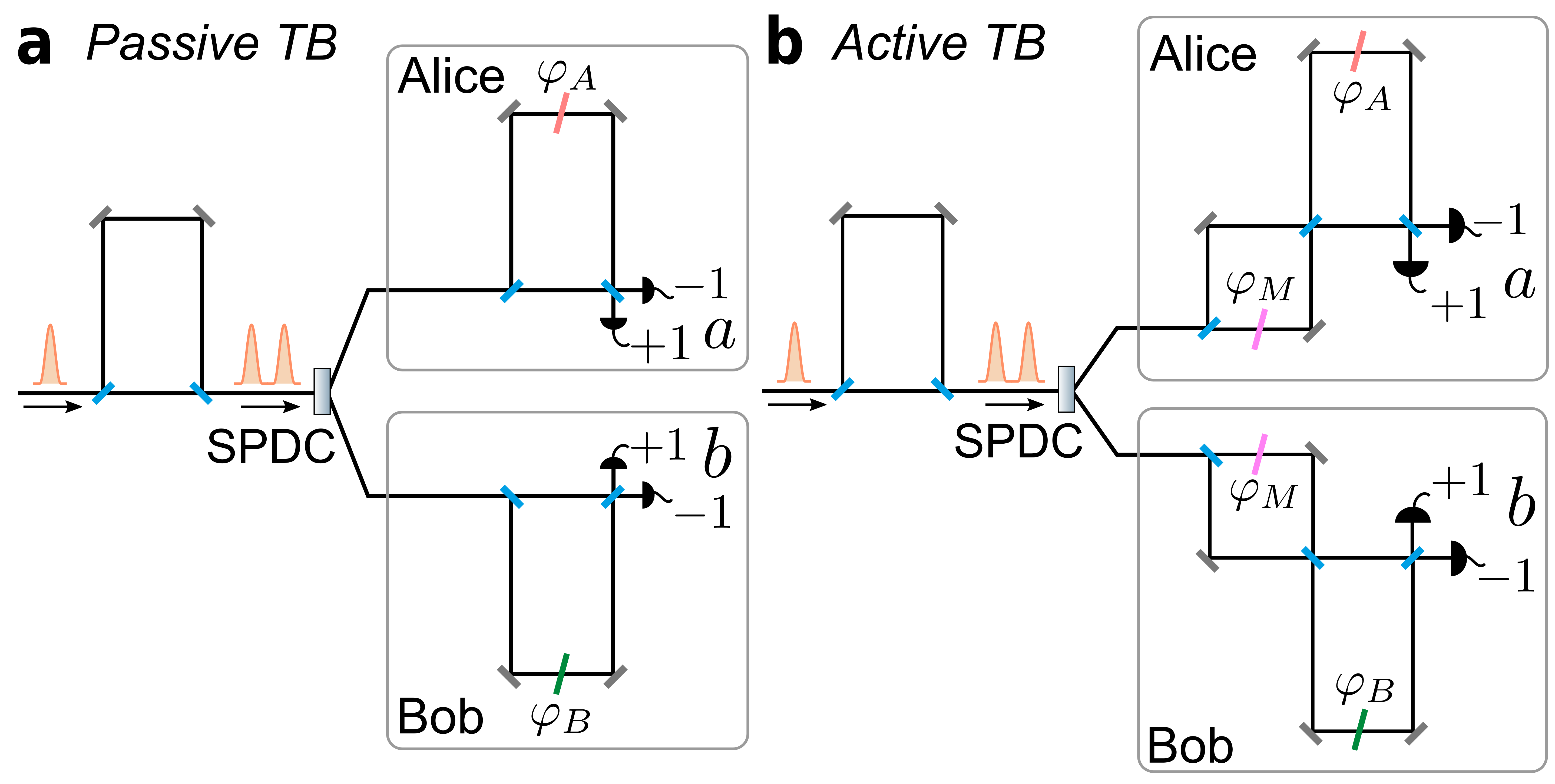}
\caption{{\it Bell-tests with time-bin entanglement.} ({\bf a})~In the passive TB, by postselecting the events detected in coincidence in the central time-slot, Alice and Bob can violate a Bell inequality, but the scheme is affected by the PSL. ({\bf b})~In the active TB, the passive beam splitter is replaced by a balanced MZI acting as an optical switch and {\red a PSL-free Bell violation is obtained} by exploiting the phase-modulator $\varphi_M$ to not discard any data.}
\label{fig_teosetup}
\end{figure}

{\it Introduction.---}In 1989, Franson conceived an interferometric setup to highlight the counter-intuitive implications of quantum mechanics~\cite{franson_1989}. He proposed to send a pair of entangled photons to two measurement stations (Alice and Bob), each composed of an unbalanced interferometer. By exploiting the quantum interference in the detection events, it should be possible to rule out local realistic mo\-dels~\cite{bell1964} by vio\-lating a Bell-CHSH inequality~\cite{CHSH_1969}. Franson's idea was first implemented by exploiting \textit{energy-time} entanglement, generated by pumping a non-linear crystal with a continuous-wave (CW) laser~\cite{Ou1990, Brendel1992,Kwiat1993}. The emitted photons are generated at the same instant, {\red which} is uncertain within the coherence time of the source, leading to indistinguishability in the alternative paths the photons will take in the measurement stations. 
Extending Franson's idea, \textit{time-bin} (TB) entanglement was introduced in 1999 by Brendel {\it et al.}~\cite{brendel1999}: the CW laser is replaced by a pulsed laser shining the non-linear crystal after passing through an unbalanced ``pump'' interferometer. Now, the photons can be emitted at two possible times, depending on the path taken by the pump-pulse in the first interferometer (Fig.~\ref{fig_teosetup}{\bf a}).
Both energy-time and time-bin entanglement have been widely used to distribute entanglement over long distances~\cite{tapster1994,tittel1998,tittel1999,marcikic2004,Inagaki2013}, and to realize fiber-based cryptographic sy\-stems~\cite{tittel2000, gisin2002}, aiming for device-independent security~\cite{Acin2006,Acin2007,Arnon-Friedman2018}, which requires the loophole-free violation of a Bell ine\-quality~\cite{Hensen2015,Giustina2015,Shalm2015,Rosenfeld2017}.

However, Aerts {\it et al.} noted that Franson's Bell-test is intrinsically affected by the \textit{postselection loophole} (PSL)~\cite{aerts1999}, which is present independently to the other loopholes (eg., locality and detection) that could affect local-realistic tests~\cite{Larsson2014}. In fact, Alice and Bob should postselect only the indistinguishable events occurring within a coincidence window $\Delta\tau_c$, discarding  
those photons arriving at different times. When performing such postselection, there exists a local-hidden-variable  (LHV) model reproducing the quantum predictions~\cite{aerts1999, jogenfors2014}. {\red Indeed,} a LHV model admits the local delays to depend on the local parameter ($\varphi_A$ or $\varphi_B$), but Alice and Bob need to compare these delays to perform the postselection. Therefore, even though the physical system is completely local, such postselection invalidates the locality assumption required to derive the  Bell's inequality. The same loophole affects the time-bin entanglement scheme shown in Fig.~\ref{fig_teosetup}{\bf a}, invalida\-ting Bell's inequality as test of local realism and enabling the hacking of Franson' scheme  when  used for cryptographic purposes~\cite{jogenfors2015}. In {\red fact}, the Bell-test gives false evidence, since the apparent violation would tell users the setup is device-independently secure, while it is {\red actually} insecure because of the PSL.

Many modifications to Franson's original scheme have been proposed to address the PSL. {\red The first one is due to Strekalov {\it et al.}~\cite{Strekalov1996}, and exploited \emph{hyper-entanglement} in polarization and energy-time to overcome the PSL by replacing the beam splitters  of Alice's and Bob's interferometers with polarizing ones. This scheme was experimentally implemented~\cite{Barreiro2005} and recently realized in an intra-city free-space link~\cite{Steinlechner2017}. 
However, requiring entanglement in both energy-time and polarization, this solution increases the experimental complexity. Afterwards,} a proposal by Cabello {\itshape et al.} modified the geometry of the interferometers by interlocking them in a \textit{hug configuration}, and introduced a \textit{local} postselection, which does not require communication between Alice and Bob~\cite{cabello2009}.  In this way, \textit{genuine} energy-time entanglement can be generated, i.e. not affected by the PSL. Soon after, table-top experiments were realized~\cite{Lima_2010,Vallone2011} as well as the distribution of genuine energy-time entanglement through 1 km of optical fibers~\cite{Cuevas2013} and its implementation in an optical fiber-network~\cite{Carvacho2015}. However, the hug configuration requires to stabilize two long interferometers whose extension is determined by
the distance between Alice and Bob: the larger the separation is, the more demanding the stabilization becomes. 
In the case of time-bin entanglement,  the original proposal mentioned the use of \emph{active} switches to prevent discarding any data~\cite{brendel1999}, such as movable mirrors synchronized with the source instead of passive beam splitters (Fig.~\ref{fig_teosetup}{\bf a}). This solution can be exploited also to overcome the PSL~\cite{jogenfors2014}, but no such scheme has been realized so far.

Here we propose and implement, for the first time, a ge\-nuine time-bin entanglement scheme allowing the violation of a Bell's inequality free of the PSL. The active switches are realized by replacing the first beam splitter, in each unbalanced interferometer of the measurement stations, with another balanced interferometer with a fast phase-shifter in one arm {\red (Fig.~\ref{fig_teosetup}{\bf b})}. By actively synchroni\-zing the phase-shifter with the pump pulses, it is possible to  use the full detection statistics, overcoming the PSL. 
The independence between Alice's and Bob's terminals, the relaxed stabilization requirements, as well as the compliance with off-the-shelves components open the possibility to exploit such scheme over long distances, 
paving the way to a conclusive loophole-free Bell-test with time-bin entanglement. {\red Indeed, the postselection procedure has long precluded the use of time-bin entanglement in loophole-free Bell experiments, performed so far by exploiting electronic spins~\cite{Hensen2015}, atoms~\cite{Rosenfeld2017} and photon polarization~\cite{Giustina2015,Shalm2015}.}

{\it Conceptual analysis of time-bin entanglement schemes.---}In the  passive time-bin scheme, a pump Mach-Zehnder interferometer (MZI) with a temporal imbalance equal to $\Delta t$ is used to split a short light pulse into two {\red(Fig.~\ref{fig_teosetup}{\bf a})}.
This light is  focused into a non-linear crystal producing photon pairs via a spontaneous parametric down conversion (SPDC) process. By optimizing the pump energy, the  generation of double photon-pairs is suppressed, and the Bell state $\ket{\Phi^+} = \left( \ket{S}_A\ket{S}_B  + \ket{L}_A\ket{L}_B \right)/\sqrt{2}$ is produced, where the indexes $A$ and $B$ represent the generated photons that are sent to Alice's and Bob's measurement stations. 
Each of these is composed of an unbalanced MZI that has the same imbalance $\Delta t$ of the pump-interferometer and can introduce a further phase shift $\varphi_A$, $\varphi_B$. The output ports of each interferometer are followed by two single-photon detectors, and the possible outcomes are labeled $a = \pm 1$ and $b = \pm 1$ for Alice and Bob respectively.

In the passive TB scheme, each photon of the pair can be detected only at three distinct times $(t_0 - \Delta t,t_0,t_0 +  \Delta t)$, due to the pump- and measurement-MZIs.
By postselecting the detection events {\red occurring} in the central time-slot, Alice realizes the projection $\{\hat{P}_{a} \}_{a=\pm 1}$ {\red with} $\hat{P}_{a} = \dyad{\psi_a}$  where 
$\ket{\psi_a} = \left(  \ket{S} + a \ \mathrm{e}^{\mathrm{i} \varphi_A}  \ket{L} \right)/\sqrt{2}$, and similar relations hold for Bob (with $a$ replaced by $b$ and $A$ by $B$).
Since  the delay is local, one could think that this should allow the violation of the Bell's inequality. There is simply no physical mechanism for the remote phase shift to influence the local delay. However, for a coincidence to occur, Bob's delay needs to coincide with Alice's, and Bob's delay is controlled by Bob's phase-shift, remotely from the point of view of Alice. This constitutes a coincidence loophole for the Bell inequality~\cite{Larsson2004}, somewhat similar to a detection loophole with 50\% detection efficiency, but much worse since it is present even when using loss-free equipment, therefore introducing an unavoidable intrinsic loophole in the setup.

Quantum mechanics provides the probabilities $\mathcal{P}_{a,b}$
for photon detections that occur  within a coincidence window $\Delta \tau_c < \Delta t$  around the central time-slot for each pair of detectors $a,b$. The probabilities
$\mathcal{P}_{a,b}$ depend on the initial state $\ket{\Phi^+}$ and on the local phase shifts $\varphi_A$, $\varphi_B$ introduced by the measurement stations and are given by $\mathcal{P}_{a,b}(\varphi_A, \varphi_B) = \frac{1}{4} \left[1 + a b \mathcal{V}  \cos(\varphi_A + \varphi_B)\right]$, where $\mathcal{V}$ is the visibility of two-photon interference. 

Disregarding the PSL, the  interference in the postselected events will seem to violate the Bell-CHSH inequality, which provides an upper limit for a combination of four correlation functions $E(\varphi_A, \varphi_B)$ with different phases $\varphi_A, \varphi_B$, when assuming the existence of a LHV model~\cite{CHSH_1969}.
The  correlation function is  
$E(\varphi_A, \varphi_B) = \sum_{a,b} a b \mathcal{P}_{a,b}(\varphi_A, \varphi_B)$
and the Bell-CHSH inequality $S \leq 2$  is given {\red by} the S-parameter
$S \equiv E(\varphi_A, \varphi_B) + E(\varphi_A', \varphi_B) + E(\varphi_A, \varphi_B') - E(\varphi_A', \varphi_B')$, 
where $\varphi_A, \varphi_A'$ and  $\varphi_B, \varphi_B'$ denote the values of the phase-shifts introduced by Alice and Bob respectively \cite{CHSH_1969}.
Quantum mechanics predicts $E^{\rm QM}(\varphi_A, \varphi_B) = \mathcal{V} \cos (\varphi_A + \varphi_B)$, {\red leading} to  a maximum S-parameter 
$S_{\rm max} = 2\sqrt{2} \mathcal{V}$ {\red if}  
$\varphi_A = -\pi/4$, $\varphi_A' = \pi/4$, $\varphi_B = 0$, and $\varphi_B'= \pi/2$. 
Hence, the Bell-CHSH inequality will seem to be violated if $\mathcal{V} > 1/\sqrt{2} \approx 0.71$. 
 
{\red Moreover,} if no postselection is applied  in the passive TB scheme, then the Bell-CHSH inequa\-lity {\red holds}, and could in principle be  violated.  However, in this case Alice implements the \textit{Positive Ope\-rator Valued Measure} (POVM)~\cite{Peres1993}  $\{\hat{\Gamma}_{a} \}_{a=\pm 1}$ with $\hat{\Gamma}_{a} = (1/4)\Id + (1/2)\hat{P}_a$,
where $\Id = \dyad{S} + \dyad{L}$ (and similarly for Bob). Thus, with no postselection, the quantum {\red detection} probabilities $\mathcal{P}_{a,b}$ lead to $S_{\rm max} = 2\sqrt{2}\mathcal{V}'$, with the overall three-peak visibility $\mathcal{V'} = \mathcal{V}/4$ and the Bell-CHSH inequality cannot be violated even with perfect visibility $\mathcal{V} = 1$.

\begin{figure}[t!]
\centering
\includegraphics[width=0.45\textwidth]{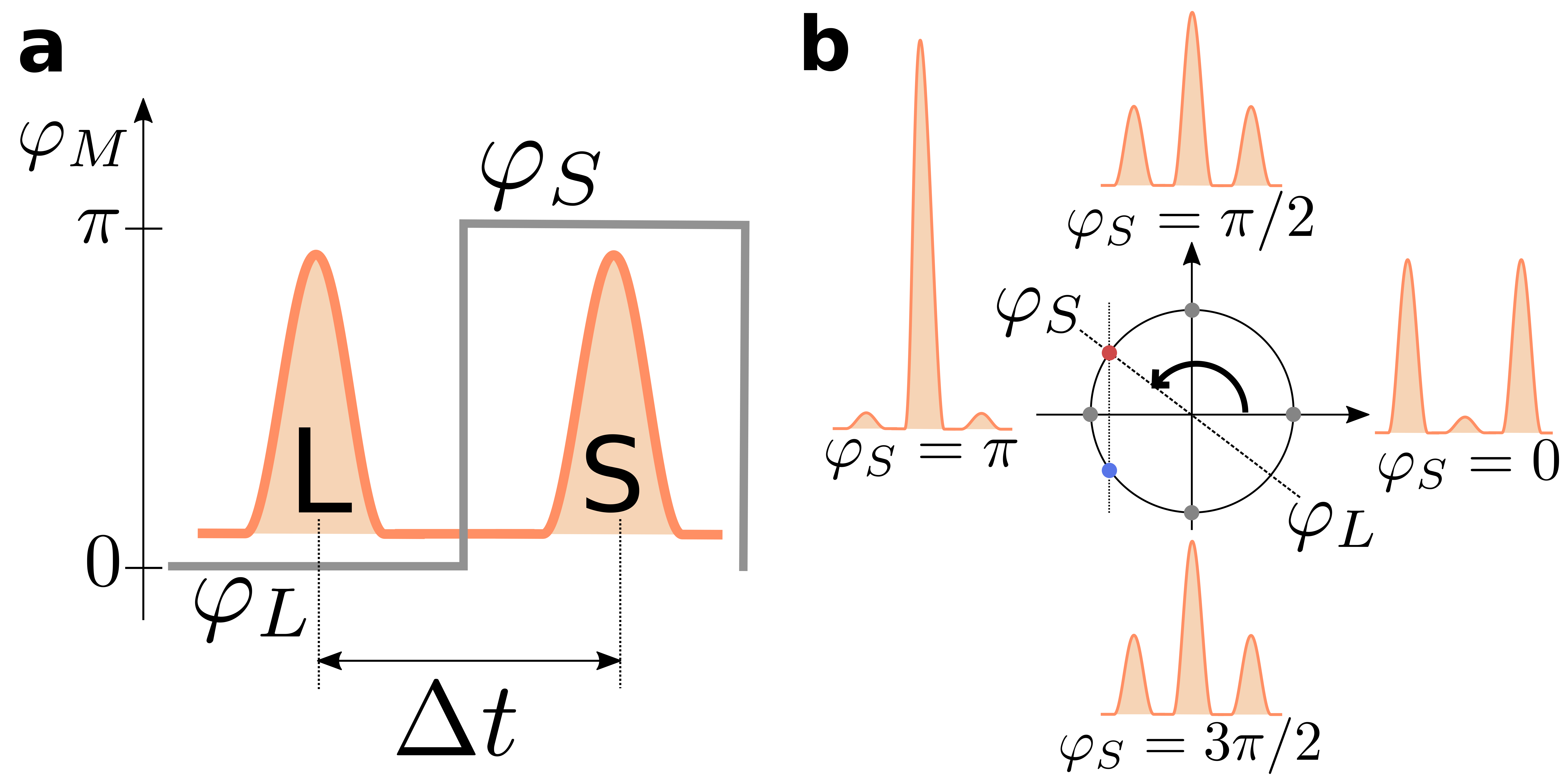}
\caption{{\it Functioning of the active TB scheme.}~({\bf a})~In a balanced MZI, the relative phase $\varphi_M$ sensed by a traveling pulse determines the output port it will exit at with probabilities $\cos^2(\varphi_M/2)$ and $\sin^2(\varphi_M/2)$. {\red A} fast modulator {\red imposes} different phase-shifts $\varphi_S$ and $\varphi_L$ to the $\ket{S}$ and $\ket{L}$ photons while they are traveling along the balanced MZI. ({\bf b})~The detection pattern depends on $\varphi_S$ and $\varphi_L=\varphi_S-\pi$. If $\varphi_S = \pi$, all detections occur in the central time-slot; if $\varphi_S = 0$ they {\red appear} only in the lateral time-slots. Any other detection histogram can be obtained with two different $\varphi_S$ values, one with $\varphi_S < \pi$ (red dot) and the other with $\varphi_S > \pi$ (blue dot). }
\label{fig_pm}
\end{figure}

Contrarily, a proper violation can be achieved in the active TB scheme here proposed (Fig.~\ref{fig_teosetup}{\bf b}). We replace the passive beam-splitter with an additional balanced MZI acting as a fast optical switch, {\red allowing} the measurement-MZI to recombine the $\ket S$ and $\ket L$ pulses, making them indistinguishable. Contrary to the passive TB scheme which recombines the two temporal modes in a probabilistic manner, our scheme deterministically compensates for the delay $\Delta t$ and  no  detections are discarded.
Indeed, by imposing the phases $\varphi_S$ and $\varphi_L=\varphi_S-\pi$ on the $\ket S$ and $\ket L$ pulses respectively, the balanced MZI determines the path they will take in the measurement-MZI,
as sketched in Fig.~\ref{fig_pm}{\bf a}.

At each detector we expect a detection pattern that depends on the value of $\varphi_S$ (Fig.~\ref{fig_pm}{\bf b}).
In our active TB scheme Alice implements the POVM $\{\hat{\Pi}_{a} \}_{a=\pm 1}$, where
$\hat{\Pi}_a = \frac{1}{2} \left(\cos^2\frac{\varphi_S}{2}\dyad{S} + 
\sin^2\frac{\varphi_L}{2}\dyad{L}\right)
+  \dyad{\chi_a}$
with
$\ket{\chi_a}=
(\mathrm{i}\mathrm{e}^{-\mathrm{i}\frac{\varphi_S}{2}}\sin\frac{\varphi_S}{2}\ket{S}+a\mathrm{e}^{\mathrm{i}(\varphi_A-\frac{\varphi_L}{2})}\cos\frac{\varphi_L}{2}\ket{L})/\sqrt{2}$.
If $\varphi_S -\varphi_L = \pi$, we have
\begin{equation}
\hat{\Pi}_a = \frac{1}{2} \cos^2\left(\frac{\varphi_S}{2}\right) \Id +  \sin^2 \left(\frac{\varphi_S}{2}\right) \hat{P}_a \ . \label{POVM_modifiedTB}
\end{equation}
{\red Setting} the phase $\varphi_S = \pi$ ({\red so} $\varphi_L = 0$), $\hat{\Pi}_a$ reduces to $\hat{P}_a$ and {\red she} projects onto the state $\ket{\psi_a}$, with no postselection procedure.
Indeed, in the detection pattern the lateral peaks ``disappear'' 
(Fig.~\ref{fig_pm}{\bf b}) and it is unnecessary to discard any data. Hence, the expected violation of Bell-CHSH inequality is PSL-free.

\begin{figure}[b!]
\centering
\includegraphics[width=0.45\textwidth]{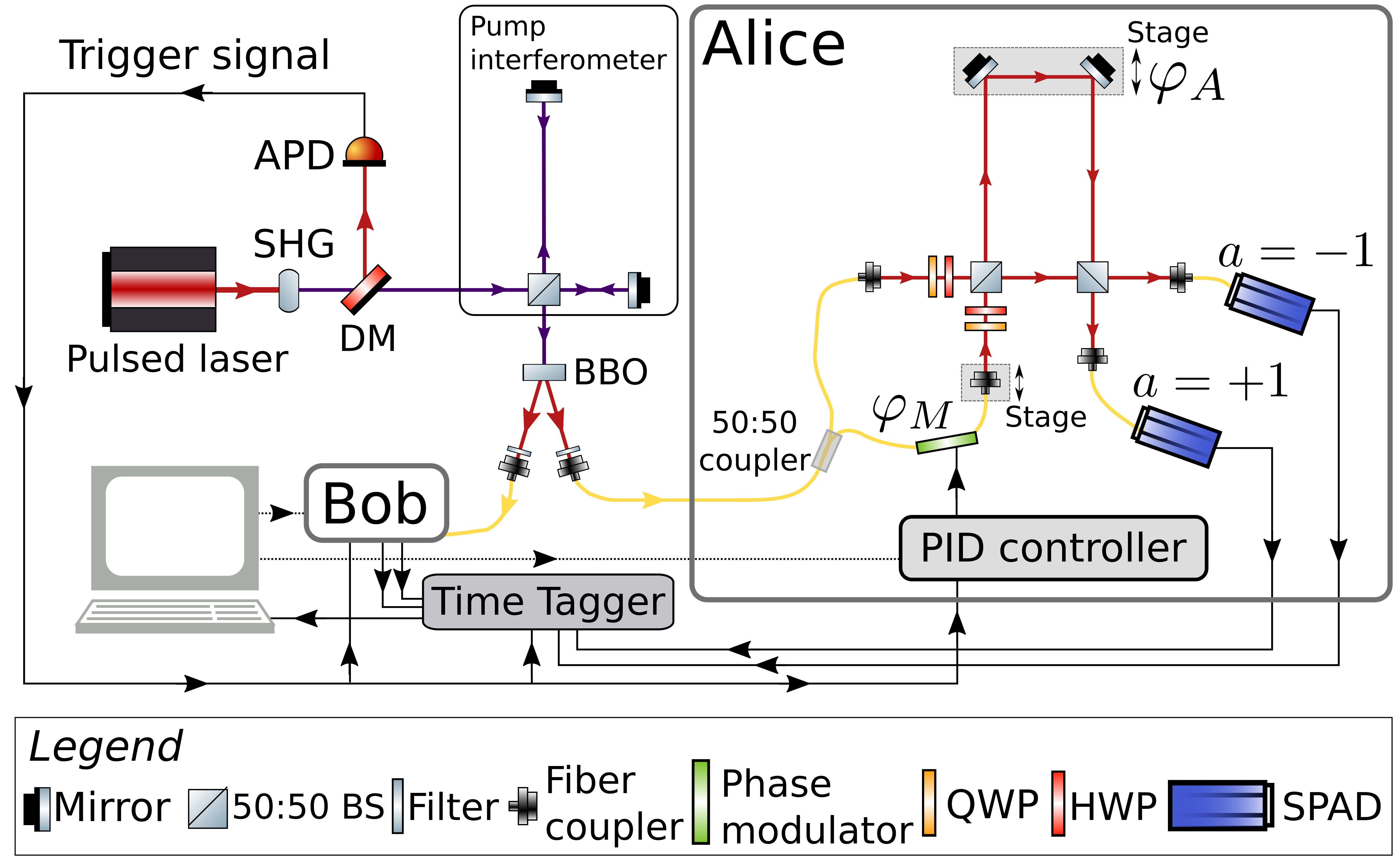}
\caption{{\it Experimental setup.} Bob's measurement station is analogue to Alice's one. APD: analog-photo-detector; DM: dichroic mirror; QWP: quarter waveplate; HWP: half waveplate.}
\label{fig_setup}
\end{figure}

{\it Description of the experiment.---}We implemented the active TB scheme proposed above by using the experimental setup sketched in Fig.~\ref{fig_setup}. 
A mode-locking laser produced a pulse train with wavelength around 808 nm, 76 MHz of repetition-rate and $\sim$150 fs of pulse-duration. This beam is used to pump a second-harmonic-generation (SHG) crystal {\red generating} coherent pulses of light up-converted to 404 nm. Each pulse passes through a free-space unbalanced Michelson interferometer (the pump-interferometer), producing a coherent state in two temporal modes. 
The imbalance between the two arms is about 90~cm, corresponding to a temporal imbalance $\Delta t \approx 3$~ns, much greater than the coherence time of the pulses. Then, the pulses pump a 2-mm long Beta-Barium Borate (BBO) crystal to produce the entangled photon state at 808 nm via type-II SPDC~\cite{kwiat1995}. 

The two photons are sent to Alice's and Bob's terminals after being  spectrally filtered (3 nm bandwidth) and collected by two single-mode optical fibers. Each station is composed of two MZIs, a balanced one and an unbalanced one. The balanced MZI is composed by a 50:50 fiber coupler which defines the two arms of the interferometer. To guarantee the zero imbalance of this MZI, a nanometric stage is placed in one of the two arms.
The balanced MZI works as a fast optical switch, since there is a fast ($\sim$GHz bandwidth) phase-modulator in one of its arms. The modulation voltage is set to $V_\pi$ such that $\varphi_S-\varphi_L=\pi$, while the DC bias of the phase-modulator is driven by an external proportional-integral-derivative (PID) controller,  that is responsible of locking the phase $\varphi_S$ to $\pi$. The complete operating principle of the PID controller is detailed in the Supplementary Material (SM). 
The two arms of the balanced MZI are recombined at a 50:50 free-space beam splitter (BS) after been optimized for polarization rotations. This BS begins the unbalanced MZI whose imbalance is equal to that of the pump-interferometer (within the coherence time $\sim$200 $\mu$s of the photons). The two mirrors of the long arm of the unbalanced MZI are placed on a nanometric piezoelectric stage to both guarantee the required imbalance $\Delta t$ and introduce the local phase shift $\varphi_A$ and $\varphi_B$ to realize the Bell-test. 
At the two output ports of the measurement stations we used two avalanche single photon detectors (SPADs, $\sim$50$\%$ detection efficiency). The detection events are then time-tagged by a time-to-digital converter (Time Tagger) with 81 ps resolution and the data are stored in a PC.

\begin{figure}[t!]
\centering
\includegraphics[width=7cm]{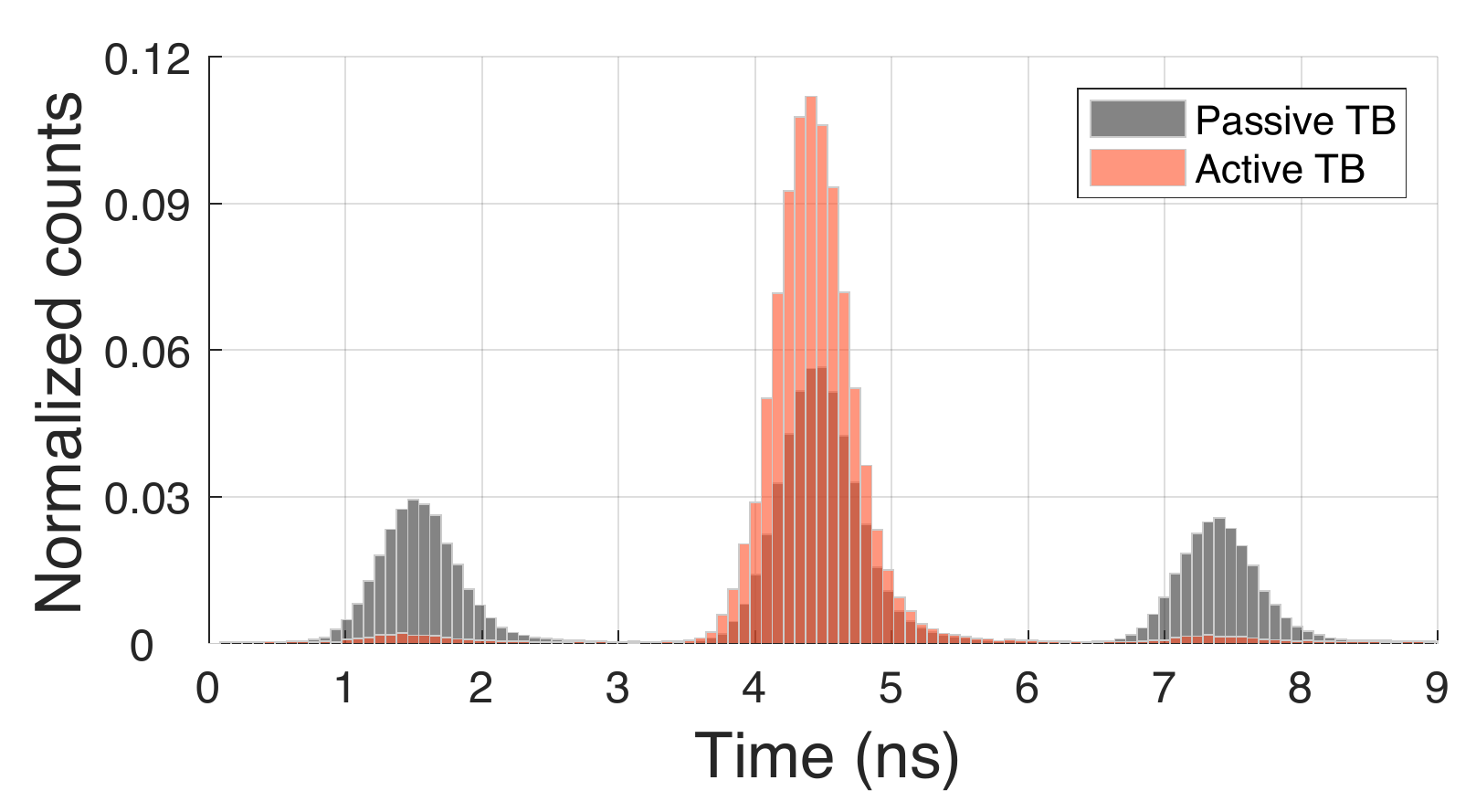}
\caption{{\red Raw detection histograms obtained with one detector for the passive TB (grey curve) and with the active TB (red curve).} The counts are normalized to fairly compare the two histograms.}
\label{fig_histo_PID}
\end{figure}

{\it Results.---}With the setup shown in Fig.~\ref{fig_setup}, we performed three different Bell-test: I) the {\it passive TB with postselection}, II) the {\it passive TB with no postselection}, III) the {\it active TB with no postselection} proposed above.
To realize I), we bypassed the balanced MZI in each of the measurement stations, hence obtaining the passive TB configuration of Fig.~\ref{fig_teosetup}{\bf a}. By choosing a coincidence window $\Delta \tau_c \approx 2.4$ ns 
and by postselecting the coincident events that occurred in the central time-slot, Alice (Bob) implemented ${\hat{P}_a}$ (${\hat{P}_b}$) and the expected Bell-CHSH violation is affected by the PSL. To realize II), we used the same configuration as in I), without discarding any data by choosing a coincidence window $\Delta \tau_c \approx 8.1$ ns, as large as the three-peak profile in the detections (Fig.~\ref{fig_histo_PID}). Now, Alice (Bob) implemented the POVM ${\hat{\Gamma}_a}$ (${\hat{\Gamma}_b}$) and no Bell-CHSH violation is expected. 
To implement III), we exploited the balanced MZI in each station and we used the PID controller to implement the phase-locking mechanism, independently at each terminal. We did not discard any data by choosing a large coincidence window as in II), and the Bell-CHSH inequality is directly applicable, since Alice (and Bob) implemented the POVM of Eq.~\eqref{POVM_modifiedTB} with $\varphi_S = \pi$. Hence, the expected Bell-CHSH violation is free of the PSL. 

A typical histogram from a detector is shown in Fig.~\ref{fig_histo_PID}, being similar the results from the other ones.
For the TB schemes I) and II), as the balanced MZI is bypassed, the expected three-peak profile ({\red grey} histogram) is found. Conversely, in our active TB scheme III), the PID controller correctly clears the lateral peaks ({\red red histogram}).

For each of the Bell-tests described above,  we first calibrated the shifts to be introduced by the nanometric stages in Alice' and Bob's unbalanced MZIs, by scanning the coincidence rate for a pair of detector by moving Bob' stage while Alice's one is fixed. From the sinusoidal pattern so obtained, we estimated the experimental visibility $\mathcal{V}_{\rm exp}$ for each scheme.  Then, we imposed the shifts $(\varphi_A, \varphi_B)$ needed to obtain the maximal violation of the Bell inequality and run the acquisition long enough for significant statistics. 

\begin{table}[b!]
\centering
\begin{tabular}{lccccc}
\hline\hline
Scheme & $\Delta\tau_c$  & PSL & $\mathcal{V}_{\rm exp}$ & $ S_{\rm exp}$ & SD \\
\hline
I) passive TB& 2.4 ns & Yes & $0.95 \pm 0.05$ & $2.58 \pm 0.03$ & $18.3$  \\ 
II) passive TB& 8.1 ns & No & $0.23 \pm 0.02$ & $0.67 \pm 0.02$ & --- \\ 
III) active TB& 8.1 ns &No & $0.89 \pm 0.03$ & $2.30 \pm 0.03$ & $9.3$\\  \hline\hline
\end{tabular}
\caption{Main results. SD refers to Standard Deviation of the Bell-CHSH violation.}
\label{table_results}
\end{table}

The results of the tests are presented in Table~\ref{table_results}. 
{\red Violations} of the Bell-CHSH inequality {\red were} obtained with the first and the third scheme with clear statistical evidence, but only {\red III)} is not affected by the PSL. The minor violation obtained is due to imperfection in the  balanced MZI alignment and in the locking procedure  during the data acquisitions needed to experimentally estimate the S-parameter $S_{\rm exp}$. {\red However,} any imperfection in the locking mechanism setting $\varphi_S =\pi$ corresponds to an effective lower visibility, but it does not introduce any loophole in the Bell-test.

{\it Conclusions.---}Time-bin encoding~\cite{brendel1999} is a resource to perform fundamental tests of quantum mechanics~\cite{Marcikic2003,Vedovato2017}, {\red quantum communications over turbulent channels~\cite{Vallone2016,Jin2018}} and to distribute entanglement over long distances~\cite{Inagaki2013}. {\red Time-bin is more robust than polarization in fiber-based implementations, since the latter suffers from polarization mode dispersion~\cite{gisin2002}, that must be carefully compensated~\cite{Hubel2007}.
Recently, a three-state Quantum Key Distribution (QKD) protocol has been proposed and implemented using both enco\-dings~\cite{Grunenfelder2018, Boaron2018}. The reduced complexity of time-bin allowed for a higher secret key rate and the longest fiber QKD link realized so far~\cite{BoaronRecord}.
Furthermore, a significant advantage of time-bin is the possibility to realize high-dimensional states (\emph{qudits}), which can provide higher QKD rates~\cite{Brougham2013,Islam2017} and that can be controlled (see the proposal~\cite{Lukens2018}) by exploiting the square phase-modulation we expe\-rimentally realized here.}

However, all time-bin entanglement realizations performed so far were affected by the PSL. {\red Another} possible way to overcome {\red it} requires to violate the ``chained'' Bell-inequalities~\cite{Braunstein1990}, but the needed visibility ($\gtrsim0.94$~\cite{jogenfors2015}) is considerably higher than the one of the Bell-CHSH inequality ($\gtrsim 0.71$). Even if such a high visibility is achievable~\cite{Tomasin2017}, our scheme strongly relaxes this requirement, since the Bell-CHSH inequality is directly applicable.  

This work is the first implementation of genuine time-bin entanglement, representing a crucial step towards its exploitation for fundamental tests of physics and the realization of the quantum internet~\cite{kimble2008}.  Our scheme can be realized using only commercial off-the-shelves fiber components and, since its stability does not depend on the distance between Alice and Bob, it is easier to be implemented with respect to the hug configuration~\cite{cabello2009}. Furthermore, as long as both the $\pi$-phase transition imposed by the modulator and the detectors jitter are shorter than the imbalance, it is possible to shorten $\Delta t$, rendering it compatible with today's photonic integrated technologies~\cite{Sibson2017, Sorianello2018}. Finally, our work makes time-bin entanglement a viable technique to obtain a loophole-free Bell violation, that is the enabling ingredient of any device-independent protocol~\cite{Acin2006,Acin2007,Arnon-Friedman2018,Acin2016}.

\newpage
\onecolumngrid

\newpage
\onecolumngrid

\section{Supplementary material}
\setcounter{equation}{0}
\renewcommand{\theequation}{S-\arabic{equation}}

\begin{figure}[h!]
\centering
\includegraphics[width=0.8\textwidth]{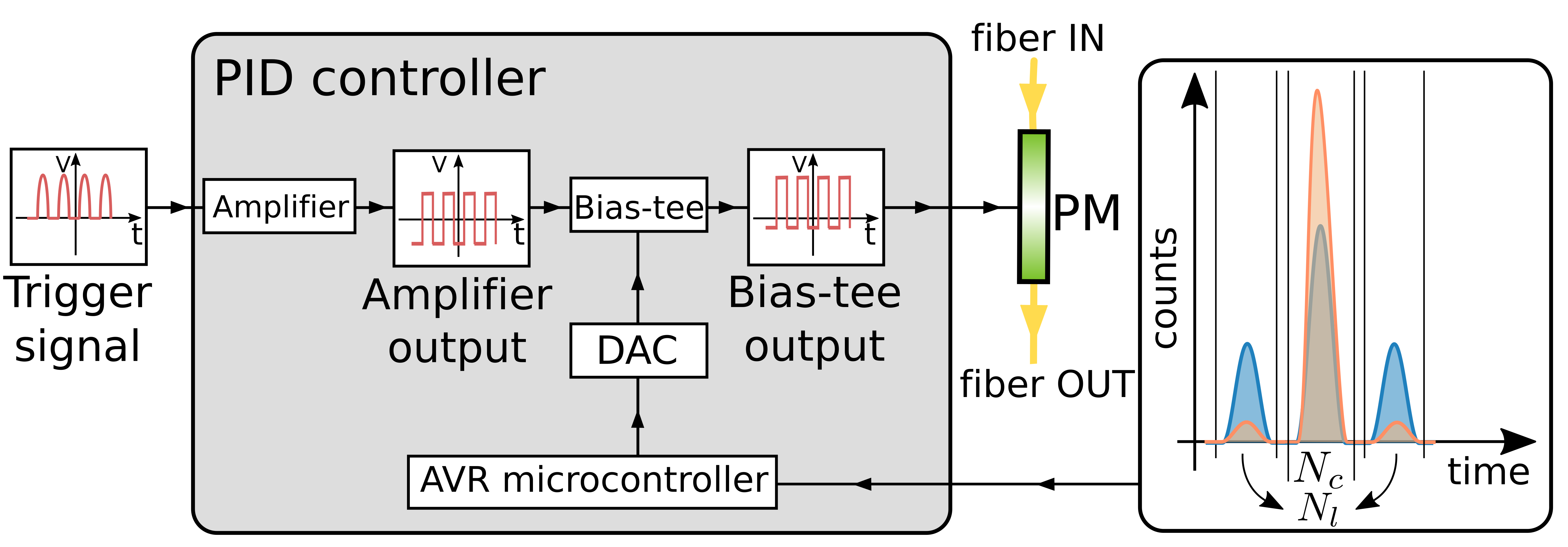}
\caption{Detailed scheme of the PID controller. }
\label{fig_dac}
\end{figure}

{\it Operating principle of the PID controller.---}In our experiment we drive the phase $\varphi_M$ introduced by the phase-modulator (PM) in the balanced MZI to make the photons take a precise path in the subsequent MZI. To realize this, we implemented the PID controller that is sketched in Fig.~\ref{fig_dac}.

First, we synchronize the phase transition with the pump-pulses that produce the photon pair.  This is performed by a fast analog-photo-diode (APD) that collects the 808 nm pulsed beam (after being separated with a dichroic mirror (DM) from the 404 nm pulse train produced by the SHG stage, see Fig.~3 of the main text) and produces an electric signal synchronized with the optical pulses. This signal is spit in two: one is then collected by the time-tagger for timing purposes and the other one is sent to the PID controller. 

The first stage of the PID controller is an amplifier (iXblue) which produces a square wave with fixed amplitude centered around 0 V. The amplitude $V_\pi$ of this wave set the strength $\Delta \varphi =  \varphi_S-\varphi_L = \pi$ of the transition introduced by the phase-modulator. The raise time of the square wave is less than 2.5 ns to guarantee that the $\pi$-transition occurs within the short-long temporal separation $\Delta t$.

The absolute value of the phase $\varphi_S$ of the balanced MZI is perturbed by temperature fluctuations and vibrations due to the environment. In order to correctly implement our scheme, we have to compensate this phase fluctuation (which occurs in the order of tens of seconds), by locking the value of $\varphi_S$ to $\pi$.

To perform this locking, the second stage of the PID controller is given by a bias-tee (MiniCircuits) which compensates the intrinsic phase shift of the balanced MZI by changing the offset voltage $V_{\rm bias}$
of the square wave produced by the amplifier. This is obtained by the combined action of an AVR micro-controller (Arduino) and a digital-to-analog  converter (DAC) by
maximizing the extinction ratio $R$ between the central and the lateral peaks $R = (N_c - N_l)/(N_c + N_l)$,
where $N_c$ are the counts associated to the central peak and $N_l$ are all the counts in the lateral ones recorded by one of the two detectors of the measurement station.  All the counts in each detector can be estimated in real-time by looking at the raw data collected by the time-tagger (QuTools), and they produce the detection histogram sketched in the inset of Fig.~\ref{fig_dac}, which corresponds to the real detection histograms presented in Fig.~4 of the main text.
 
To successfully lock $\varphi_S$ to $\pi$ the PID controller has to first 
evaluate its real-time value by observing the detection histogram and computing $R$. Unfortunately, there is no one-to-one correspondence between the extinction ratio and the phase $\varphi_S$. Indeed, for each possible value of $R$ there exist two possible values for $\varphi_S$ that reproduce the observed histograms (with the exception of $0$ and $\pi$), as shown in Fig.~2b of the main text. Therefore, we must include an additional information that allows us to distinguish between the two possible phase values. This information is given by the derivative of the extinction ratio. If an increase of the phase value causes an increase of the ratio, we choose the phase $0<\varphi_S<\pi$ (requiring further increase to reach $\pi$). Otherwise, we choose the phase $\pi<\varphi_S<2\pi$ (requiring a decrease to reach $\pi$). Since the PID requires an error function that is equal to zero when the objective is reached, we choose the function $E_{\varphi_S} = \sgn\left(\frac{d R}{d\varphi_S} \right) \frac{N_l}{N_c}$, which guarantees that the PID's objective is both to lock the value of $\varphi_S$ to $\pi$ and to identify correctly the value of the phase, since the symmetry between the two possible phase values is broken by the sign of the derivative of the extinction ratio.

\end{document}